\documentclass{article}

\usepackage[T1]{fontenc}
\usepackage{microtype}
\usepackage{booktabs}
\usepackage{tabularx}
\usepackage{hyperref}

\usepackage[preprint]{icml2026}

\usepackage{amsmath,amssymb}
\usepackage{listings}
\usepackage{xcolor}
\usepackage{xspace}

\newif\ifblind
\blindfalse

\newcommand{\blind}[2]{\ifblind#1\else#2\fi}
\newcommand{\anonurl}{\textsc{[url omitted]}\xspace}
\newcommand{\anonpkg}{\textsc{[package omitted]}\xspace}
\newcommand{\anonname}{\textsc{[organization omitted]}\xspace}

\newcommand{\axleurl}{\blind{\anonurl}{\url{https://axle.axiommath.ai}}}
\newcommand{\axledocsurl}{\blind{\anonurl}{\url{https://axle.axiommath.ai/v1/docs/}}}
\newcommand{\axlemcpurl}{\blind{\anonurl}{\url{https://mcp.axiommath.ai}}}
\newcommand{\axleexamplesurl}{\blind{\anonurl}{\url{https://colab.research.google.com/github/AxiomMath/axiom-lean-engine/blob/main/examples/starting_demo.ipynb}}}

\newcommand{\axlepypi}{\blind{\anonpkg}{\texttt{axiom-axle}}}
\newcommand{\axlemcppypi}{\blind{\anonpkg}{\texttt{axiom-axle-mcp}}}

\newcommand{\axiommathname}{\blind{\anonname}{Axiom Math}}

\newcommand{\axle}{\textsc{Axle}\xspace}
\newcommand{\axlefull}{Axiom Lean Engine\xspace}
\newcommand{\lean}{\textsc{Lean}\xspace}
\newcommand{\leanfour}{\textsc{Lean~4}\xspace}
\newcommand{\mathlib}{\textsc{Mathlib}\xspace}

\newcommand{\numtools}{14\xspace}

\newcommand{\tool}[1]{\texttt{#1}}
\newcommand{\verifyproof}{\tool{verify\_proof}\xspace}
\newcommand{\extractdecls}{\tool{extract\_decls}\xspace}
\newcommand{\merge}{\tool{merge}\xspace}
\newcommand{\normalize}{\tool{normalize}\xspace}
\newcommand{\repairproofs}{\tool{repair\_proofs}\xspace}
\newcommand{\simplifytheorems}{\tool{simplify\_theorems}\xspace}
\newcommand{\disprove}{\tool{disprove}\xspace}
\newcommand{\havetolemma}{\tool{have2lemma}\xspace}
\newcommand{\sorrytolemma}{\tool{sorry2lemma}\xspace}
\newcommand{\theoremtosorry}{\tool{theorem2sorry}\xspace}

\newcommand{\totalrequests}{500 million\xspace}

\newcommand{\K}[1]{\textbf{#1}}

\icmltitlerunning{\axle: A Cloud Infrastructure for \leanfour Theorem Proving Utilities}

\begin{document}

\twocolumn[
\icmltitle{\axle: A Cloud Infrastructure for \leanfour Theorem Proving Utilities}

\icmlsetsymbol{equal}{*}
\icmlsetsymbol{prior}{\textdagger}

\begin{icmlauthorlist}
\icmlauthor{Jimmy Xin}{axiom}
\icmlauthor{Alex Schneidman}{tml,prior}
\icmlauthor{Chris Cummins}{rsi,prior}
\icmlauthor{Karun Ram}{axiom}
\icmlauthor{Srihari Ganesh}{axiom}
\icmlauthor{Jannis Limperg}{axiom}
\end{icmlauthorlist}

\icmlaffiliation{axiom}{Axiom Math}
\icmlaffiliation{tml}{Thinking Machines Lab}
\icmlaffiliation{rsi}{Recursive Superintelligence, Inc}

\icmlcorrespondingauthor{Jimmy Xin}{axle@axiommath.ai}

\icmlkeywords{Lean 4, theorem proving, formal verification, AI for mathematics, cloud infrastructure}

\vskip 0.3in
]

\printAffiliationsAndNotice{\textsuperscript{\textdagger}Work done while at Axiom Math.}

\begin{abstract}
We present \axle (\axlefull), a cloud service for \leanfour proof manipulation, extraction, and verification. Recent progress in AI for mathematics---reinforcement learning pipelines, agentic proving workflows, dataset curation---demands \leanfour tooling that scales to millions of requests while remaining correct and robust; existing infrastructure offers parallel compilation but not scalable proof verification, higher-level proof manipulation, multi-version support, or per-request isolation at the throughput modern AI workflows require. \axle provides \numtools \leanfour metaprogramming tools spanning strict proof verification, declaration metadata extraction, semantic source manipulation, lemma extraction, and deterministic proof repair and simplification. The service runs as a multi-tenant cloud deployment with per-request isolation and concurrent support for multiple \leanfour and \mathlib versions, accessible via a Python SDK, command-line interface, web UI, MCP server, and HTTP API. \axle is publicly available and free to use at \axleurl~and via the \axlepypi~PyPI package, with no local \leanfour installation required. It has served over \totalrequests requests to date\blind{ and is the underlying infrastructure for several formal-mathematics projects [specific applications omitted for double-blind review].}{ and is the underlying infrastructure for \axiommathname's proving efforts, including its 12/12 score on the 2025 Putnam competition.}
\end{abstract}

\section{Introduction}

\leanfour~\cite{lean4} is a programming language and proof assistant in which users write programs and proofs together, with correctness guaranteed by a dependent type system.
Its community-driven mathematics library, \mathlib~\cite{mathlib}, contains over 250{,}000 formal theorems and has become a central resource for formalization.
\leanfour has also emerged as the language of choice for AI-driven mathematics: AlphaProof~\cite{alphaproof}, DeepSeek-Prover~\cite{deepseekprover}, Kimina-Prover~\cite{kiminaprover}, Seed-Prover~\cite{seedprover}, Aristotle~\cite{aristotle}, Leanstral~\cite{leanstral}, and more all use \leanfour as their formal backend.
These efforts---spanning model training, reinforcement-learning pipelines, agentic proving workflows, and dataset curation---share a common need for \leanfour tooling that is not only fast and correct, but that supports the high concurrency these workloads require.

Several prior systems provide infrastructure for interacting with \leanfour at scale (\S\ref{sec:relatedwork}). However, these efforts have various limitations in scope. First, compilation is not verification: A passing compile accepts proofs containing \texttt{sorry}, unsound axioms, or incorrectly restated theorems---all common failure modes in AI-generated proofs. Strict verification techniques exist~\cite{lean4checker,comparator,safeverify}, but no prior system makes them available as a fast, scalable cloud service. Second, prior work lacks additional higher-level manipulation primitives and metadata extraction that agentic and dataset-curation workloads depend on, all of which require custom functionality implemented in \lean's metaprogramming framework. Third, shared-process architectures risk state leakage and crash propagation across requests. Fourth, most infrastructure is pinned to a single \leanfour or \mathlib version, whereas AI systems frequently target multiple versions for training data from different eras or for cross-version compatibility. Finally, existing systems do not offer elastic scalability under a simple interface: automatically scaling capacity up during training runs and down during idle periods, with load balanced across workers.

\axle (\axlefull) was built to address these gaps. It provides \numtools \leanfour metaprogramming tools, served from a hosted multi-tenant cloud deployment with per-request isolation and concurrent support for multiple \leanfour versions. \axle has been deployed in production\blind{; specific deployment examples are omitted for double-blind review.}{, including as the underlying infrastructure for \axiommathname's 12/12 score on the 2025 Putnam competition and autoformalizations of Fel's conjecture on numerical semigroups~\cite{fels} and the partial-regularity-of-primes result related to Vandiver's conjecture~\cite{partialvandiver}, among others.} The service is free and publicly accessible; Appendix~\ref{app:resources} collects the relevant resource URLs, installation commands, and support channels.

Our contributions are:

\begin{enumerate}
    \item \textbf{Complete suite of proof-manipulation tools.} \axle provides \numtools metaprogramming tools, each implemented as a \leanfour metaprogram: strict proof verification, declaration metadata extraction, semantic source manipulation, lemma extraction, and deterministic proof repair and simplification.
    \item \textbf{Strict proof verification at scale.} \axle's \verifyproof rejects candidates containing \texttt{sorry}, non-whitelisted axioms, or signature mismatches---failure modes that are common in AI-generated proofs and that plain compilation accepts. \axle makes strict \leanfour proof verification available at production throughput rather than as a single-process command-line tool.
    \item \textbf{Multi-environment support.} A single \axle endpoint serves multiple \leanfour and \mathlib versions concurrently, enabling cross-version compatibility checks and seamless support for models targeting different \leanfour versions.
    \item \textbf{Free public cloud service.} \axle is deployed as a free, publicly accessible cloud service with per-request isolation and automatic retries; users access it through a web UI at \axleurl, a Python SDK and CLI via the \axlepypi~PyPI package, an MCP server, and an HTTP API, none of which require a local \leanfour installation.
\end{enumerate}

\section{Motivating Applications}
\label{sec:applications}

\axle serves the proof-engineering needs of AI-driven \leanfour mathematics. Although the tools and infrastructure are general, the principal workloads fall into three categories: reinforcement learning and model training, agentic proving, and dataset curation. Each places distinct demands on \axle's tools, latency profile, and scalability; subsequent sections describe the tools and infrastructure that meet them.

\subsection{Reinforcement Learning and Model Training}

Reinforcement learning over formal proofs has become a central paradigm for training \leanfour proving models~\cite{alphaproof,deepseekprover,kiminaprover,seedprover,aristotle}.
A typical RL step samples thousands of candidate proofs from a policy, evaluates each, and uses the outcome as a reward signal. This workload bursts at each rollout boundary into a flood of concurrent verification requests, and demands strict correctness: A verifier that accepts \texttt{sorry}, unsound axioms, or restated theorems would teach the policy to game the reward (see Appendix~\ref{app:verify-examples}). Per-request isolation is equally important, since a malformed candidate must not corrupt subsequent verifications on the same worker. \axle's strict-verification tool \verifyproof (\S\ref{sec:tools}) and its per-request isolation directly target this regime, and the service handles the concurrent verification load these rollouts produce. Raw speed matters too: A typical verification completes well under a second, keeping step-level rollout times bounded even with thousands of parallel candidates. Importantly, \verifyproof adds little overhead beyond plain compilation, so the strictness checks do not penalize throughput.

\subsection{Agentic Proving}

A second class of workloads is \emph{agentic} pipelines~\cite{copra,apollo,hilbert,gauss,numinaleanagent}, in which an LLM iteratively proposes, inspects, and refines proof candidates.
A single attempt may involve dozens of turns: Propose a proof, verify it, inspect errors or remaining goals, extract those goals as standalone lemmas, attempt them recursively, repair localized failures, and merge the resulting fragments into a complete proof. These workflows demand a richer toolkit than verification alone---lemma extraction from \texttt{sorry} positions and elaboration errors, declaration metadata and dependency analysis for retrieving relevant context, merging of partial proofs from parallel attempts, and deterministic repair of common errors in LLM-generated drafts. Sitting inside the agent's inner loop, the tools must be fast, robust, and stable; they must be invoked dozens of times per attempt without bottlenecking the agent or producing spurious errors that are uninterpretable to an LLM. Appendix~\ref{app:demo} demonstrates one such decompose-solve-merge loop end-to-end on a non-trivial \leanfour lemma.

\subsection{Dataset Curation and Filtering}

A third class of workloads is the construction and filtering of training datasets from existing \leanfour corpora such as \mathlib. Common operations include extracting theorem-level metadata in bulk, filtering by structural properties (dependency complexity, proof length, tactic usage), deduplicating across multiple sources, and normalizing or simplifying statements for downstream consumption. These workloads are throughput-bound rather than latency-bound---a single build may issue millions of requests over hours or days---and can span multiple \leanfour and \mathlib versions, since training corpora are often pinned to specific releases or built across version boundaries. \axle's high request throughput, concurrent multi-version support, and metadata- and dependency-extraction tools are aimed at this workload.

\section{Service Overview}
\label{sec:architecture}

\axle is a hosted, multi-tenant cloud service. A client issues a tool request---a tool name, an input payload, and a target \leanfour environment---to \axle's HTTPS endpoint, and receives a JSON response with the tool's structured output. Authentication is by API key. No local \leanfour installation is required.

\subsection{Client Surfaces}

Clients reach \axle through any of five interchangeable methods for submitting JSON-encoded tool calls: a Python SDK and a command-line interface, both distributed via the \axlepypi~PyPI package; an MCP server distributed via \axlemcppypi~(or hosted at \axlemcpurl) for agent integrations; a hosted web UI at \axleurl; and the underlying HTTPS API.

\subsection{Multi-Version Environments}

Each request carries an \texttt{environment} field that selects a particular \leanfour version paired with a \mathlib snapshot and any project-specific pre-built dependencies. A single \axle deployment serves multiple environments concurrently behind the same endpoint, which AI workloads use for cross-version compatibility checks, evaluation against legacy proofs, and serving models trained against different \mathlib eras. The available environments are exposed by the API; by default, the public tier serves a range of recent \leanfour--\mathlib release snapshots, with additional custom environments available on request.

\subsection{Per-Request Isolation}

Each request runs in its own sandboxed process. State a request mutates---loaded definitions, set options, registered attributes, allocated memory---does not persist into any later request, and a crash or runaway elaboration in one request does not affect concurrent or subsequent requests. The sandbox additionally blocks network access and prevents the candidate from writing to the filesystem---useful for AI-generated input that may contain unbounded loops or unintended side effects.

\subsection{Scaling and Reliability}

\axle scales in response to sustained load and uses queueing to absorb short request bursts from existing capacity. Per-request wall-clock timeouts are configurable by the caller, up to a service-wide cap; transient infrastructure failures (worker crashes, network errors) trigger automatic retries without surfacing to the client. Per-API-key fair-share queueing ensures one heavy caller cannot starve other tenants on the deployment.

\section{Tools}
\label{sec:tools}

All of \axle's tools are implemented as \lean metaprograms operating on parsed and elaborated \lean syntax. This section walks through the principal tools; a complete summary is given in Table~\ref{tab:tools}, and per-tool reference documentation is available online.\footnote{\axledocsurl}

\subsection{Compilation Checking and Proof Verification}

The simplest tool is \tool{check}, which compiles a \lean file and reports any errors, warnings, and linter messages. Its strict counterpart, \verifyproof, takes a candidate file and a separate \emph{formal statement} file defining the theorems to be proved, and accepts the candidate only if its declarations match the formal statements and constitute a valid proof.

Compilation alone is frequently unsuitable for AI-generated proof workloads. \leanfour permits user-defined macros, custom tactics, metaprograms, and new axioms; the same features that make \lean expressive also create many ways for a file to compile without proving the intended statement. The \verifyproof tool rejects any declarations that:
\begin{itemize}
    \item contain \texttt{sorry};
    \item use any axiom outside the small whitelist of known-consistent axioms shipped in the \lean standard library (\texttt{propext}, \texttt{Quot.sound}, \texttt{Classical.choice});
    \item have a type which does not match the formal statement (e.g., the candidate proves a weakened or restated version of the theorem);
    \item are marked \texttt{unsafe}, and can therefore use kernel-bypassing primitives such as \texttt{unsafeCast}.
\end{itemize}

\axle intentionally does not defend against certain rare classes of adversarial inputs, prioritizing performance over maximal strictness. Several existing tools provide stronger forms of strict proof verification, including \texttt{lean4checker}~\cite{lean4checker}, Comparator~\cite{comparator}, and SafeVerify~\cite{safeverify}. For scalability, \verifyproof assumes that every declaration in the loaded \lean environment was added through \lean's normal kernel-checked elaboration path: it does not re-verify the environment from scratch and hence does not defend against inputs that use \lean metaprogramming to install unchecked declarations directly into the environment and make invalid proofs appear valid. For the typical case of AI-generated proofs this is rarely an issue (\S\ref{sec:eval-compare}), and \verifyproof is substantially faster and more scalable than the alternatives, as Section~\ref{sec:evaluation} discusses. Appendix~\ref{app:verify-examples} gives concrete examples for this discussion.

\subsection{Declaration Extraction}

\extractdecls extracts structured metadata for every declaration in a file---theorems, definitions, structures, instances, and so on. Its principal output for each declaration is a self-contained \lean snippet that includes the declaration together with all its transitive dependencies within the file, ready to compile in isolation. Alongside this, \extractdecls returns per-declaration metadata: name, kind, signature, body, source-line range, a stable type hash, whether the declaration contains \texttt{sorry}, lightweight statistics such as proof length and tactic counts, any \lean-emitted messages, and dependency information at three levels: \emph{type} (constants appearing in the stated type), \emph{value} (constants in the elaborated body, including intermediate constants used by the elaborator during proof construction), and \emph{syntactic} (constants the source code names literally).
Dataset pipelines use \extractdecls to turn a multi-declaration file into one standalone snippet per declaration; agents use it to isolate the chunk of a larger file they need to work on.

\subsection{Semantic Merge}

\merge is the inverse of \extractdecls: It takes a collection of \lean source fragments and combines them into a single compilable file. Na\"ive concatenation does not work in general---declarations must appear in dependency order, duplicate declarations must be unified, and name conflicts must be resolved.

\merge handles each. Duplicate declarations are detected by $\alpha$-equivalence of types, optionally using definitional equality. Among duplicates, \merge keeps a single compiling representative---that also has only compiling dependencies---if one exists; otherwise it keeps one invalid representative. Surviving declarations are then topologically sorted by their dependencies.
When two declarations share a name but their types do not match, the conflict is resolved by renaming one of them and updating references using \axle's \tool{rename} tool.

A typical application is combining proof fragments from multiple AI candidates into a single compiling file---a routine operation in agentic workflows.

\subsection{File- and Theorem-Level Transformations}

\axle exposes a family of source-level transformations. Among these are \tool{rename}, which renames a declaration and propagates the change to every reference site---handling namespace resolution, dot notation, constructor projections, and named structure fields---and \normalize, which expands abbreviated names, unfolds notation, and removes \texttt{section} and \texttt{namespace} blocks while preserving compile semantics like maintaining \texttt{noncomputable} modifiers from a removed section. Other transformations include \tool{theorem2lemma}, which swaps the \texttt{theorem} keyword for \texttt{lemma} (and vice versa), and \theoremtosorry, which replaces proofs with \texttt{sorry} stubs while leaving statements intact. Several of these (for example, \theoremtosorry and \normalize) are routine building blocks in dataset-preparation pipelines: \theoremtosorry turns finished proofs into sorry-stubbed problem statements, and \normalize standardizes surface syntax across corpora before training.

\subsection{Lemma Extraction}

\havetolemma and \sorrytolemma lift proof obligations from inside a partial proof into standalone top-level lemmas. The mechanism is built on \mathlib's \texttt{extract\_goal} tactic:
At the target location---an unresolved \texttt{have} for \havetolemma, an explicit \texttt{sorry} or the position of an elaboration error for \sorrytolemma---the tool captures the proof state, generalizes over its local context, and emits a fresh lemma whose statement equals the captured goal.

The extracted lemma is callsite-reconstructable: \axle can additionally rewrite the original proof to invoke the new lemma in place of the original obligation, with the correct arguments supplied. A separate transformation tool, \tool{have2sorry}, replaces \texttt{have} steps with \texttt{sorry} stubs without extracting a lemma; it is useful for sketching proof outlines that subsequent calls fill in. An agent stuck on a proof step uses \havetolemma or \sorrytolemma to isolate the obligation, solves the resulting lemma independently, and optionally reassembles the pieces with \merge.

We note that due to bugs and limitations in existing utilities like \texttt{extract\_goal}, this tool, and several others, may fail on certain inputs. More details on the failure modes can be found in the documentation.

\subsection{Proof Optimization}

\simplifytheorems and \repairproofs transform existing proofs through pipelines of small, deterministic passes---\simplifytheorems to tidy up a compiling proof, \repairproofs to coax a failing one into compiling.

\simplifytheorems prunes structure the proof does not actually use, like tactics flagged as unused or unreachable by \lean's linters, and \texttt{have} steps whose hypothesis is never referenced. The pipeline iterates to a fixed point, with each pass surfacing new opportunities for the next; if any pass introduces an error, the iteration is rolled back, so the tool never breaks a working proof.

\repairproofs targets common failure modes in AI-generated drafts by attempting to prove residual unproven goals, truncating proofs when there are no goals remaining, attempting to replace unsafe constructs that \verifyproof would reject (like the \texttt{native\_decide} tactic), and cleaning up cross-version porting issues such as \texttt{set\_option} commands naming options that no longer exist.

\begin{table}[t]
\centering
\small
\begin{tabularx}{\linewidth}{@{}lX@{}}
\toprule
\textbf{Tool} & \textbf{Output} \\
\midrule
\tool{check}           & errors, warnings, linter messages \\
\verifyproof           & verification pass/fail, failed declarations \\
\extractdecls          & per-declaration metadata, dependencies, standalone snippets \\
\merge                 & merged compilable file with conflicts resolved \\
\normalize             & normalized source preserving semantics \\
\tool{rename}          & file with renames applied and references updated \\
\repairproofs          & repaired file with unfixable goals left as-is \\
\simplifytheorems      & file with unused tactics and hypotheses pruned \\
\disprove              & counterexample or proof of negation \\
\theoremtosorry        & file with proofs replaced by \texttt{sorry} \\
\tool{theorem2lemma}   & file with \texttt{theorem} swapped for \texttt{lemma} \\
\havetolemma           & top-level lemmas extracted from \texttt{have} positions \\
\tool{have2sorry}      & file with \texttt{have} steps replaced by \texttt{sorry} \\
\sorrytolemma          & top-level lemmas extracted from \texttt{sorry} positions \\
\bottomrule
\end{tabularx}
\caption{Complete list of \axle's tools and their primary outputs.}
\label{tab:tools}
\end{table}

\section{Evaluation}
\label{sec:evaluation}

We evaluate two aspects of \axle's design: per-request cost and aggregate throughput against alternative ways to run \lean at scale (\S\ref{sec:eval-perf}), and strict-proof-verification quality and speed against other published strict checkers (\S\ref{sec:eval-compare}). The throughput workload is drawn from the public \texttt{banach1729/goedel-workbook-lean427} dataset~\cite{goedel427}, a set of Lean Workbook competition-math proofs pinned to \lean~4.27.0. The strict-checker comparison uses an internal paired \verifyproof corpus drawn from production traffic. To keep the candidate set identical across systems on the throughput workload, every request's import block is normalized to a single \texttt{import Mathlib}; candidates with non-Mathlib dependencies are dropped (discussed as a limitation in \S\ref{sec:conclusion}). Additional experiments are in Appendix~\ref{app:eval-ablations}, with setup details in \S\ref{app:eval-setup}.

\subsection{Latency and Throughput on a Single Machine}
\label{sec:eval-perf}

Table~\ref{tab:perf-main} compares \axle to two alternatives on a randomly-sampled $5{,}000$-proof \tool{check} workload from the Goedel workbook dataset: Kimina Lean Server~\cite{kimina}---a long-running \lean REPL pool---and a baseline that spawns a fresh \lean process per request. All three run at concurrency~$8$ on a single \texttt{r7a.2xlarge} AWS instance (8~vCPU, 64~GiB), with a 150~s per-request budget.

\begin{table*}[t]
\centering
\small
\begin{tabular}{l r r r r r}
\toprule
system & median latency & p90 latency & throughput & total wall & verdict counts \\
\midrule
\axle                                                & 1.05 s  & 5.96 s  & 2.09 req/s          & 2389 s & 4708 / 287 / 5 \\
Kimina Lean Server~\cite{kimina}                     & 0.75 s  & 6.38 s  & \textbf{2.13 req/s} & 2351 s & 4708 / 287 / 5 \\
\lean invoked directly (no shared \mathlib)  & 5.14 s  & 9.81 s  & 1.03 req/s          & 4873 s & 4708 / 287 / 5 \\
\bottomrule
\end{tabular}
\caption{Per-request latency and aggregate throughput on a $5{,}000$-proof \tool{check} workload from the Goedel workbook dataset, at concurrency~$8$ on a single \texttt{r7a.2xlarge}. ``Verdict counts'' is pass / fail / timeout at the 150~s budget. All three systems return the same verdict on every one of the $5{,}000$ proofs.}
\label{tab:perf-main}
\end{table*}

All three systems return the same verdict on every one of the $5{,}000$ proofs (Table~\ref{tab:perf-main}). On throughput, raw \lean is about $2{\times}$ slower than the other two: it pays the full cold-start cost on each request of reloading \mathlib modules, which dominates when the median proof needs less than a second of actual elaboration. \axle and Kimina both keep \mathlib preloaded and avoid paying the start-up cost. Kimina is marginally faster ($0.75$~s vs.\ $1.05$~s latency, $2.13$~req/s vs. $2.09$~req/s throughput). This gap comes from \axle's per-request isolation (\S\ref{sec:architecture}): every request runs in a sandboxed worker process, which adds overhead, while Kimina dispatches into a warm REPL from a long-running pool. Appendix~\ref{app:eval-axle-concurrency} reports an \axle-only concurrency scaling experiment on the same workload.

\subsection{\verifyproof vs.\ Strict Proof Checkers}
\label{sec:eval-compare}

\verifyproof is one of several published tools that aim to perform strict proof checking in \leanfour: Given a candidate and a formal statement, decide whether the candidate proves the statement. We compare it against Comparator~\cite{comparator} and SafeVerify~\cite{safeverify} on a set of $1{,}000$ production \verifyproof requests, running each alternative on the same \texttt{r7a.4xlarge} (16~vCPU, 128~GiB) at concurrency~$4$. The per-request budget is $600$~s.

\K{Speed.} Table~\ref{tab:cmp-speed} reports the per-request median latency, aggregate throughput, total wall time, and success rate. \verifyproof's median is sub-second, roughly $10{\times}$ faster than SafeVerify and $99{\times}$ faster than Comparator at the median. The throughput gap is dominated by each alternative's per-request defense work. In addition to reloading imports on every request, both Comparator and SafeVerify run an \textit{environment replay} step that re-typechecks each declaration to verify its soundness, and they include additional safeguards such as sandboxing, rebuilding expressions, and detecting suspicious use of \texttt{Nat} literals. \verifyproof trusts that constants were added through the standard kernel-checked elaboration path (\S\ref{sec:tools}, Appendix~\ref{app:verify-examples}) and so does not re-check the environment. \verifyproof finishes the workload at $0.43$~req/s, while SafeVerify runs at $0.107$~req/s and Comparator at $0.026$~req/s.

\begin{table*}[t]
\centering
\small
\begin{tabular}{l r r r r}
\toprule
system & median latency & throughput & total wall & success \\
\midrule
\axle \verifyproof                & 0.97 s  & \textbf{0.43 req/s} & 9{,}205 s          & \textbf{100.0\%} \\
SafeVerify~\cite{safeverify}      & 10.1 s  & 0.107 req/s         & 37{,}258 s         & 99.9\% \\
Comparator~\cite{comparator}      & 95.7 s  & 0.026 req/s         & 151{,}267 s        & 93.8\% \\
\bottomrule
\end{tabular}
\caption{Per-request median latency and aggregate throughput on $1{,}000$ paired production \verifyproof requests; all three systems use the same instance at the same concurrency (Appendix~\ref{app:eval-setup}). ``Success'' is the share of requests that produce an accept-or-reject verdict within the $600$~s per-request budget; the remaining are heavier candidates that exhaust the budget.}
\label{tab:cmp-speed}
\end{table*}

\K{Agreement.} \axle and Comparator agree on every one of the $938$ conclusive pairs (Table~\ref{tab:cmp-agreement}). The remaining $62$ are Comparator-side timeouts at the $600$~s budget: Comparator's median latency is $95.7$~s and its slowest candidates cross $10$ minutes. \axle and SafeVerify agree on $992$ of $999$ conclusive pairs. The $7$ disagreements all arise from a single mechanism: When a candidate places \texttt{private} helpers inside a namespace, \lean stores each private declaration under a mangled symbol that includes the surrounding module name, and the same source compiled into two different module names ends up with two distinct sets of mangled symbols. SafeVerify performs a whole-environment comparison between the target and submission \texttt{.olean} files and flags every such mangled symbol as missing. However, the user-named theorem itself matches in both files, and so \verifyproof and Comparator both accept the submission. Appendix~\ref{app:case-tool-diffs} shows the pattern with an example.

None of the disagreements reflect a soundness issue on \verifyproof's part. We did not encounter naturally-occurring requests that incorrectly pass \verifyproof while failing Comparator or SafeVerify, although such examples can be deliberately crafted (\S\ref{app:case-vp-misses}).

\begin{table*}[t]
\centering
\small
\setlength{\tabcolsep}{5pt}
\begin{tabular}{l r r r r r r}
\toprule
comparison & both accept & both reject & vp stricter & alt.\ stricter & inconclusive & agreement \\
\midrule
\verifyproof vs.\ Comparator    & 58 & 880 & 0 & 0  & 62  & \textbf{100.00\%} \\
\verifyproof vs.\ SafeVerify    & 51 & 941 & 0 & 7  & 1   & \textbf{99.30\%} \\
\bottomrule
\end{tabular}
\caption{Verdict agreement on $1{,}000$ paired production \verifyproof requests. ``vp stricter'' indicates that \verifyproof rejected and the alternative accepted; ``alt. stricter'' indicates the reverse. ``Inconclusive'' counts requests where one side produced no clean verdict (excluded from the agreement percentage).}
\label{tab:cmp-agreement}
\end{table*}

\section{Related Work}
\label{sec:relatedwork}

\subsection{\leanfour Services and Tooling}

The Lean REPL~\cite{leanrepl} provides JSON-over-stdio access to a long-running \lean process. LeanInteract~\cite{leaninteract} wraps it as a Python library for parallel processing; leanclient~\cite{leanclient} and lean-lsp-mcp\footnote{\url{https://github.com/oOo0oOo/lean-lsp-mcp}} take a different route and wrap \lean's LSP, the latter packaged as an MCP server for agentic proof sessions. The Kimina Lean Server~\cite{kimina} pools many REPL instances behind a REST API with header-level LRU caching, aimed at high-throughput verification in RL training. It matches \axle in throughput. However, none of these provide a scalable hosted service, isolation between requests, multi-version support, or anything beyond compilation and tactic-state access.

LeanDojo~\cite{leandojo} pairs a Python interface for proof-state interaction with an extracted dataset and a retrieval-augmented prover; Pantograph~\cite{pantograph} exposes \leanfour's proof states for tactic-level interaction and tree-search procedures; ProofWala~\cite{proofwala} provides a multilingual Coq/Lean framework for proof-step data collection and parallel proof search. These operate at the tactic-state level, requiring a long-running proof-state object \axle deliberately does not maintain; \axle works at the declaration level (file in, transformed source or per-declaration metadata out) and is request-scoped and stateless, which is what enables the tool suite (\S\ref{sec:tools}) and per-request isolation. Jixia~\cite{jixia} is a command-line static-analysis tool that extracts source-level metadata---declarations, symbols, elaboration info, and the parsed AST---from a compiled \leanfour file through a plugin architecture; it is closest in scope to a single \axle tool (\extractdecls).

\subsection{Strict Proof Verification}

lean4checker~\cite{lean4checker}, Comparator~\cite{comparator}, and SafeVerify~\cite{safeverify} run candidates in more scrutinous settings than \verifyproof, catching the environment-manipulation attacks of Appendix~\ref{app:verify-examples} at the cost of speed and concurrency. \axle's deliberate trade-off is throughput on cooperating-client populations; the alternatives suit verification of untrusted external code. Note that lean4checker inspects a candidate environment but does not verify proofs against an external \textit{formal statement}, so it is not included in the comparison in \S\ref{sec:eval-compare}.

\subsection{Agentic Proving Pipelines}

APOLLO~\cite{apollo} and HILBERT~\cite{hilbert} are agentic systems that combine LLM reasoning with formal verification through decompose-repair-recombine loops. They exercise exactly the operations \axle exposes (sub-lemma extraction, source repair, semantic merging), but implement them as heuristic string- or model-based orchestrations around the Lean REPL rather than as \lean metaprograms behind a shared service.

\section{Conclusion}
\label{sec:conclusion}

\axle is a cloud service for \leanfour proof manipulation, built on the premise that these tools belong in consolidated infrastructure for AI4math rather than remaining afterthoughts in each downstream system. Its \numtools \leanfour metaprograms---spanning capabilities in verification, source manipulation, dependency and lemma extraction, simplification, and repair---serve all three principal AI4math workloads (reinforcement learning of prover models, agentic proving pipelines, and scalable dataset curation) from a single multi-tenant service with concurrent multi-version support.

\subsection{Limitations}

\axle has two limitations worth noting.

\paragraph{Multi-file projects.} Custom \lean projects can be added as additional environments (Section~\ref{sec:architecture}), but onboarding them is higher-friction than selecting a new \lean version: The project must be built and registered ahead of time, with no way to drop in arbitrary user-supplied dependencies at request time. As a result, the public service does not offer generalized multi-file support. Each environment exposes a fixed import header, typically just \texttt{import Mathlib}, and external imports beyond what that header provides are not accepted.

\paragraph{Proof search and interactive sessions.} \axle is built around request-scoped, stateless tool invocations and does not support proof-tree search, interactive proof sessions, or tactic-state-level manipulations---all of which require a long-running proof-state object \axle does not maintain. Users who need that style of interaction should look at tools like Pantograph~\cite{pantograph} and lean-lsp-mcp.

\subsection{Future Work}

Concrete directions include broader multi-file support beyond fixed-header environments, detailed proof state extraction, and additional capabilities for tools like \simplifytheorems and \repairproofs. We welcome community suggestions for additional tools and features.

\section*{Acknowledgements}
We thank the Lean FRO for their help and feedback, the rest of the \axiommathname~team as \axle's initial internal users, and our early public users for their feedback during the initial release.

\section*{Impact Statement}

\axle is freely available infrastructure for AI-driven \leanfour formal mathematics. We see no specific societal risks beyond those already attendant on advances in machine learning for theorem proving.

\paragraph{AI Disclosure} AI tools were used for prose drafting and editing throughout. All design decisions and quantitative results are the authors'.

\bibliography{technical_report}

\begin{thebibliography}{27}
\providecommand{\natexlab}[1]{#1}
\providecommand{\url}[1]{\texttt{#1}}
\expandafter\ifx\csname urlstyle\endcsname\relax
  \providecommand{\doi}[1]{doi: #1}\else
  \providecommand{\doi}{doi: \begingroup \urlstyle{rm}\Url}\fi

\bibitem[Aniva et~al.(2025)Aniva, Sun, Miranda, Barrett, and
  Koyejo]{pantograph}
Aniva, L., Sun, C., Miranda, B., Barrett, C., and Koyejo, S.
\newblock {Pantograph}: A machine-to-machine interaction interface for advanced
  theorem proving, high-level reasoning, and data extraction in {Lean} 4.
\newblock In Gurfinkel, A. and Heule, M. (eds.), \emph{Tools and Algorithms for
  the Construction and Analysis of Systems (TACAS)}, pp.\  104--123. Springer
  Nature Switzerland, 2025.
\newblock \doi{10.1007/978-3-031-90643-5_6}.

\bibitem[{banach1729}(2026)]{goedel427}
{banach1729}.
\newblock Goedel-workbook-lean-4.27: {Lean}{}~workbook competition-math proofs
  from {DeepSeek-Prover-V1.5}, migrated to {Lean}{}~4.27.0 /
  {Mathlib}{}~v4.27.0.
\newblock
  \url{https://huggingface.co/datasets/banach1729/goedel-workbook-lean427},
  2026.

\bibitem[Chen et~al.(2026{\natexlab{a}})Chen, Cummins, Eltschig,
  et~al.]{partialvandiver}
Chen, E., Cummins, C., Eltschig, B., et~al.
\newblock Almost all primes are partially regular.
\newblock \emph{arXiv preprint arXiv:2602.05090}, 2026{\natexlab{a}}.

\bibitem[Chen et~al.(2026{\natexlab{b}})Chen, Cummins, Grubisic, et~al.]{fels}
Chen, E., Cummins, C., Grubisic, D., et~al.
\newblock Fel's conjecture on syzygies of numerical semigroups.
\newblock \emph{arXiv preprint arXiv:2602.03716}, 2026{\natexlab{b}}.

\bibitem[Chen et~al.(2025)]{seedprover}
Chen, L. et~al.
\newblock {Seed-Prover}: Deep and broad reasoning for automated theorem
  proving.
\newblock \emph{arXiv preprint arXiv:2507.23726}, 2025.

\bibitem[de~Moura \& Ullrich(2021)de~Moura and Ullrich]{lean4}
de~Moura, L. and Ullrich, S.
\newblock The {Lean} 4 theorem prover and programming language.
\newblock In \emph{Automated Deduction (CADE 28)}, volume 12699 of
  \emph{Lecture Notes in Computer Science}, pp.\  625--635. Springer, 2021.
\newblock \doi{10.1007/978-3-030-79876-5_37}.

\bibitem[Dos~Santos et~al.(2025)Dos~Santos, de~Sax{\'e}, Wang, Wang,
  Bak\v{s}ys, {\"U}nsal, Liu, Liu, and Li]{kimina}
Dos~Santos, M., de~Sax{\'e}, H., Wang, H., Wang, R., Bak\v{s}ys, M., {\"U}nsal,
  M., Liu, J., Liu, Z., and Li, J.
\newblock {Kimina Lean Server}: A high-performance {Lean} server for
  large-scale verification.
\newblock In \emph{NeurIPS 2025 Workshop on AI for Math}, 2025.
\newblock arXiv preprint arXiv:2504.21230.

\bibitem[Dressler(2025)]{leanclient}
Dressler, O.
\newblock leanclient: A {Python} client for interacting with the {Lean} 4
  language server.
\newblock \url{https://github.com/oOo0oOo/leanclient}, 2025.

\bibitem[{FrenzyMath}(2024)]{jixia}
{FrenzyMath}.
\newblock jixia: A static analysis tool for {Lean} 4.
\newblock \url{https://github.com/frenzymath/jixia}, 2024.

\bibitem[{GasStationManager}(2025)]{safeverify}
{GasStationManager}.
\newblock Safeverify: A {Lean} 4 script for robustly verifying submitted proofs
  of theorems and implementations of functions.
\newblock \url{https://github.com/GasStationManager/SafeVerify}, 2025.

\bibitem[Hubert et~al.(2025)]{alphaproof}
Hubert, T. et~al.
\newblock Olympiad-level formal mathematical reasoning with reinforcement
  learning.
\newblock \emph{Nature}, 2025.
\newblock \doi{10.1038/s41586-025-09833-y}.

\bibitem[{Lean community}(2023)]{leanrepl}
{Lean community}.
\newblock A read--eval--print loop for {Lean} 4.
\newblock \url{https://github.com/leanprover-community/repl}, 2023.

\bibitem[{Lean FRO}(2023)]{lean4checker}
{Lean FRO}.
\newblock lean4checker: Replay the {Environment} for a given {Lean} module.
\newblock \url{https://github.com/leanprover/lean4checker}, 2023.
\newblock Integrated into Lean 4 v4.28.0 as {\tt leanchecker}; repository
  archived 2026.

\bibitem[{Lean FRO}(2025)]{comparator}
{Lean FRO}.
\newblock Comparator: A trustworthy judge for {Lean} proofs.
\newblock \url{https://github.com/leanprover/comparator}, 2025.

\bibitem[Liu et~al.(2026)Liu, Zhou, Zhu, Dos~Santos, He, Liu, Wang, Xie, Zhao,
  Wang, Zhi, Li, and Li]{numinaleanagent}
Liu, J., Zhou, Z., Zhu, Z., Dos~Santos, M., He, W., Liu, J., Wang, R., Xie, Y.,
  Zhao, J., Wang, Q., Zhi, L., Li, J., and Li, W.
\newblock {Numina-Lean-Agent}: An open and general agentic reasoning system for
  formal mathematics.
\newblock \emph{arXiv preprint arXiv:2601.14027}, 2026.

\bibitem[{Math Inc.}(2025)]{gauss}
{Math Inc.}
\newblock {Gauss}: An agent for autoformalization.
\newblock \url{https://www.math.inc/gauss}, 2025.

\bibitem[{Mistral AI}(2026)]{leanstral}
{Mistral AI}.
\newblock {Leanstral}: An open-source code agent for {Lean} 4.
\newblock \url{https://mistral.ai/news/leanstral}, 2026.

\bibitem[Ospanov et~al.(2025)Ospanov, Farnia, and Yousefzadeh]{apollo}
Ospanov, A., Farnia, F., and Yousefzadeh, R.
\newblock {APOLLO}: Automated {LLM} and {Lean} collaboration for advanced
  formal reasoning.
\newblock \emph{arXiv preprint arXiv:2505.05758}, 2025.

\bibitem[Poiroux et~al.(2025)Poiroux, Kun{\v{c}}ak, and Bosselut]{leaninteract}
Poiroux, A., Kun{\v{c}}ak, V., and Bosselut, A.
\newblock {LeanInteract}: A {Python} interface for {Lean} 4.
\newblock \url{https://github.com/augustepoiroux/LeanInteract}, 2025.

\bibitem[Ren et~al.(2025)Ren, Shao, Song, Xin, Wang, Zhao, Zhang, Fu, Zhu,
  Yang, et~al.]{deepseekprover}
Ren, Z.~Z., Shao, Z., Song, J., Xin, H., Wang, H., Zhao, W., Zhang, L., Fu, Z.,
  Zhu, Q., Yang, D., et~al.
\newblock {DeepSeek-Prover-V2}: Advancing formal mathematical reasoning via
  reinforcement learning for subgoal decomposition.
\newblock \emph{arXiv preprint arXiv:2504.21801}, 2025.

\bibitem[Thakur et~al.(2024)Thakur, Tsoukalas, Wen, Xin, and Chaudhuri]{copra}
Thakur, A., Tsoukalas, G., Wen, Y., Xin, J., and Chaudhuri, S.
\newblock An in-context learning agent for formal theorem-proving.
\newblock In \emph{Conference on Language Modeling (COLM)}, 2024.
\newblock arXiv preprint arXiv:2310.04353.

\bibitem[Thakur et~al.(2025)Thakur, Tsoukalas, Durrett, and
  Chaudhuri]{proofwala}
Thakur, A., Tsoukalas, G., Durrett, G., and Chaudhuri, S.
\newblock {ProofWala}: Multilingual proof data synthesis and theorem-proving.
\newblock \emph{arXiv preprint arXiv:2502.04671}, 2025.

\bibitem[{The Harmonic Team}(2025)]{aristotle}
{The Harmonic Team}.
\newblock {Aristotle}: {IMO}-level automated theorem proving.
\newblock \emph{arXiv preprint arXiv:2510.01346}, 2025.

\bibitem[{The mathlib Community}(2020)]{mathlib}
{The mathlib Community}.
\newblock The {Lean} mathematical library.
\newblock In \emph{Proceedings of the 9th ACM SIGPLAN International Conference
  on Certified Programs and Proofs (CPP)}, pp.\  367--381, 2020.
\newblock \doi{10.1145/3372885.3373824}.

\bibitem[Varambally et~al.(2025)Varambally, Voice, Sun, Chen, Yu, and
  Ye]{hilbert}
Varambally, S., Voice, T., Sun, Y., Chen, Z., Yu, R., and Ye, K.
\newblock {Hilbert}: Recursively building formal proofs with informal
  reasoning.
\newblock \emph{arXiv preprint arXiv:2509.22819}, 2025.

\bibitem[Wang et~al.(2025)]{kiminaprover}
Wang, H. et~al.
\newblock {Kimina-Prover} {Preview}: Towards large formal reasoning models with
  reinforcement learning.
\newblock \emph{arXiv preprint arXiv:2504.11354}, 2025.

\bibitem[Yang et~al.(2023)Yang, Swope, Gu, Chalamala, Song, Yu, Godil, Prenger,
  and Anandkumar]{leandojo}
Yang, K., Swope, A., Gu, A., Chalamala, R., Song, P., Yu, S., Godil, S.,
  Prenger, R., and Anandkumar, A.
\newblock {LeanDojo}: Theorem proving with retrieval-augmented language models.
\newblock In \emph{Advances in Neural Information Processing Systems
  (NeurIPS)}, 2023.

\end{thebibliography}
\bibliographystyle{icml2026}

\newpage
\appendix
\onecolumn

\section{Further Evaluation}
\label{app:eval-ablations}

\subsection{Experimental Setup}
\label{app:eval-setup}

\paragraph{Hardware and isolation.} Every measurement runs on a single AWS EC2 instance, with the specific instance varying across experiments. All systems are pre-warmed before measurement with a single \texttt{import Mathlib} request. The per-request budget is $150$~s for the \tool{check} workload (\S\ref{sec:eval-perf}) and $600$~s for the \verifyproof workload (\S\ref{sec:eval-compare}). For \axle, the \lean executor workflow is isolated from the client-interaction logic (rate-limiting, queueing, load balancing, etc.). For a fair comparison with raw \lean invocations and Kimina, measurements only include per-request computation. Public users should expect additional (minimal) server-side and network overhead.

\paragraph{Workloads.} The latency/throughput comparison (\S\ref{sec:eval-perf}) uses $5{,}000$ random proofs sampled with a fixed seed from the public \texttt{banach1729/goedel-workbook-lean427} HuggingFace dataset~\cite{goedel427}---DeepSeek-Prover-V1.5 outputs on the Lean Workbook competition-math corpus, published as compilable \lean files migrated to \lean~4.27.0. Every system and concurrency setting sees the same $5{,}000$ in the same order. Imports are normalised to a single \texttt{import Mathlib} and candidates with non-Mathlib dependencies are dropped (discussed in \S\ref{sec:conclusion}). The dataset's published claim is that $94.1\%$ of its proofs compile cleanly on the target snapshot; our sample lands at $94.16\%$ ($4708$ / $5000$) under all three systems, which is consistent with the reported figure. The strict-checker comparison (\S\ref{sec:eval-compare}) uses a separate $1{,}000$-request paired \verifyproof sample drawn randomly from production traffic; each row pairs a target formal statement with a candidate proof from the same production request.

\paragraph{Latency/throughput setup (\S\ref{sec:eval-perf}).} \axle, Kimina Lean Server, and the direct-invocation baseline all run on a single \texttt{r7a.2xlarge} (8~vCPU, 64~GiB), each at concurrency~$8$. Production \axle's \tool{check} does some extra bookkeeping Kimina's \texttt{/verify} does not---custom per-declaration logging and sorry-state extraction---so for the comparison we disable that bookkeeping in \axle, leaving the same parse-and-elaborate work Kimina does. Kimina is built from source against \lean~4.27.0 and runs with its default settings. The direct-invocation baseline runs separate \lean 4.27.0 subprocesses, one per request.

\paragraph{Strict-checker setup (\S\ref{sec:eval-compare}).} \axle, Comparator, and SafeVerify all run on a single \texttt{r7a.4xlarge} (16~vCPU, 128~GiB) at concurrency~$4$. Each alternative is driven through a thin Python wrapper that compiles target and submission to oleans (for SafeVerify) or stages them in a shared Lake project (for Comparator), then runs the upstream binary. The per-request budget is $600$~s for every system; Comparator's median replay is $95.7$~s but several take up to $10$ minutes, and 62 requests exceed the time budget. \axle is configured with \texttt{use\_def\_eq=False} on the per-target signature check, matching what Comparator and SafeVerify do (neither reduces expressions before comparing).

\subsection{Concurrency Scaling for \axle on the Same Instance}
\label{app:eval-axle-concurrency}

To show how \axle scales by number of in-flight requests, we sweep its concurrency on the same $5{,}000$-proof Goedel-Lean4.27 workload and the same \texttt{r7a.2xlarge} as \S\ref{sec:eval-perf}.

\begin{table}[t]
\centering
\small
\begin{tabular}{r r r r r}
\toprule
in-flight & median latency & p90 latency & throughput & scaling \\
\midrule
1   & 0.96 s & 5.73 s & 0.29 req/s & $1.00{\times}$ \\
2   & 0.97 s & 5.80 s & 0.57 req/s & $1.98{\times}$ \\
4   & 0.98 s & 5.79 s & 1.11 req/s & $3.87{\times}$ \\
\textbf{8}   & \textbf{1.05 s} & \textbf{5.99 s} & \textbf{2.09 req/s} & $\mathbf{7.27{\times}}$ \\
16  & 4.68 s & 11.4 s & 2.09 req/s & $7.27{\times}$ \\
\bottomrule
\end{tabular}
\caption{\axle's per-request median latency, p90 latency, and aggregate throughput at varying in-flight concurrency on a \texttt{r7a.2xlarge} ($8$~vCPU), on the $5{,}000$-proof Goedel-Lean4.27 \tool{check} workload. ``Scaling'' is throughput relative to ${c{=}1}$. Throughput scales linearly to the box's vCPU count; pushing beyond $8$ concurrent requests oversubscribes the physical cores, inflating latency without adding throughput.}
\label{tab:perf-conc}
\end{table}

As expected, throughput scales near-linearly from ${c{=}1}$ through ${c{=}8}$. Median and p90 latency remain flat in the same range, confirming that adding workers does not slow individual requests; it only adds capacity. Past ${c{=}8}$ the workers oversubscribe the $8$ physical cores: throughput plateaus at $2.09$~req/s while median latency jumps roughly $4.5{\times}$ (to $4.68$~s) and p90 nearly doubles. The verdict split is identical at every concurrency ($4708$ / $287$ / $5$); concurrency changes the speed, not the verdict.

\section{Tool Demo}
\label{app:demo}

This appendix walks through using \axle to prove a non-trivial \leanfour lemma. Taken from the public \texttt{starting\_demo} notebook in the \axle examples repository,\footnote{\axleexamplesurl} it illustrates the \emph{subproblem decomposition} pattern: Starting from an LLM-generated proof attempt that does not compile end-to-end, we use \axle to break it into independent sub-obligations, solve each, and reassemble the result.

\lstdefinestyle{axledemo}{
  basicstyle=\ttfamily\footnotesize,
  breaklines=true,
  breakatwhitespace=false,
  columns=fullflexible,
  keepspaces=true,
  showstringspaces=false,
  extendedchars=true,
  inputencoding=utf8,
  frame=single,
  framerule=0.2pt,
  framesep=4pt,
  xleftmargin=4pt,
  xrightmargin=4pt,
  mathescape=false,
  literate=
    {ℕ}{{$\mathbb{N}$}}1
    {ℤ}{{$\mathbb{Z}$}}1
    {∣}{{$\mid$}}1
    {∀}{{$\forall$}}1
    {∃}{{$\exists$}}1
    {⊢}{{$\vdash$}}1
    {←}{{$\leftarrow$}}1
    {→}{{$\rightarrow$}}1
    {↑}{{$\uparrow$}}1
    {·}{{$\cdot$}}1
    {⟨}{{$\langle$}}1
    {⟩}{{$\rangle$}}1
    {≤}{{$\leq$}}1
    {≥}{{$\geq$}}1
    {≠}{{$\neq$}}1
    {α}{{$\alpha$}}1
    {β}{{$\beta$}}1
    {λ}{{$\lambda$}}1
    {≡}{{$\equiv$}}1
    {⇒}{{$\Rightarrow$}}1
    {«}{{\guillemotleft}}1
    {»}{{\guillemotright}}1
    {√}{{$\sqrt{\phantom{x}}$}}1
    {ℝ}{{$\mathbb{R}$}}1
}

\subsection*{The target statement}

We aim to prove a divisibility lemma on greatest common divisors of consecutive odd numbers.

\begin{lstlisting}[style=axledemo,language={}]
import Mathlib

lemma gcd_dvd_two_mul_diff (m n : ℕ) :
    (Nat.gcd (2 * m + 1) (2 * n + 1) : ℤ) ∣ 2 * ((m : ℤ) - (n : ℤ)) := by sorry
\end{lstlisting}

\subsection*{The initial LLM attempt}

The agent produces the following proof attempt---a reasonable two-step strategy, but with tactics that do not all elaborate cleanly:

\begin{lstlisting}[style=axledemo,language={}]
lemma gcd_dvd_two_mul_diff (m n : ℕ) :
    (Nat.gcd (2 * m + 1) (2 * n + 1) : ℤ) ∣ 2 * ((m : ℤ) - (n : ℤ)) := by
  have h1 : Nat.gcd (2 * m + 1) (2 * n + 1) ∣ (2 * m + 1) - (2 * n + 1) := by
    apply Nat.gcd_dvd_sub
  have h2 : (2 * m + 1) - (2 * n + 1) = 2 * (m - n) := by ring
  rw [h2] at h1
  rcases le_or_gt n m with hn | hn
  · rw [← Nat.cast_sub hn]; exact_mod_cast h1
  · rw [show (2 : ℤ) * (↑m - ↑n) = -(↑(2 * n + 1) - ↑(2 * m + 1)) from by push_cast; ring]
    exact dvd_neg.mpr (dvd_sub (by exact_mod_cast Nat.gcd_dvd_right _ _)
      (by exact_mod_cast Nat.gcd_dvd_left _ _))
\end{lstlisting}

Calling \tool{check} on the attempt confirms it fails. The relevant errors:

\begin{lstlisting}[style=axledemo,language={}]
error: Unknown constant `Nat.gcd_dvd_sub`
error: unsolved goals
  m n : ℕ
  h1 : (2 * m + 1).gcd (2 * n + 1) ∣ 2 * m + 1 - (2 * n + 1)
  ⊢ 1 + m * 2 - (1 + n * 2) = (m - n) * 2
\end{lstlisting}

Rather than discarding the attempt, we use \axle to isolate the failing pieces and fix each independently.

\subsection*{Step 1: Decompose with \havetolemma}

\havetolemma lifts each \texttt{have} step into a standalone lemma and rewrites the main proof to invoke it at the call site, preserving the overall structure:

\begin{lstlisting}[style=axledemo,language=Python]
result = await client.have2lemma(
    content=initial_attempt,
    reconstruct_callsite=True,
    environment=ENVIRONMENT,
)
decomposed = result.content
\end{lstlisting}

The decomposed file:

\begin{lstlisting}[style=axledemo,language={}]
import Mathlib

lemma gcd_dvd_two_mul_diff.h1 (m n : ℕ) :
    (2 * m + 1).gcd (2 * n + 1) ∣ 2 * m + 1 - (2 * n + 1) := sorry

lemma gcd_dvd_two_mul_diff.h2 (m n : ℕ)
    (h1 : (2 * m + 1).gcd (2 * n + 1) ∣ 2 * m + 1 - (2 * n + 1)) :
    2 * m + 1 - (2 * n + 1) = 2 * (m - n) := sorry

lemma gcd_dvd_two_mul_diff (m n : ℕ) :
    (Nat.gcd (2 * m + 1) (2 * n + 1) : ℤ) ∣ 2 * ((m : ℤ) - (n : ℤ)) := by
  have h1 : Nat.gcd (2 * m + 1) (2 * n + 1) ∣ (2 * m + 1) - (2 * n + 1) :=
    gcd_dvd_two_mul_diff.h1 m n
  have h2 : (2 * m + 1) - (2 * n + 1) = 2 * (m - n) :=
    gcd_dvd_two_mul_diff.h2 m n h1
  rw [h2] at h1
  rcases le_or_gt n m with hn | hn
  · rw [← Nat.cast_sub hn]; exact_mod_cast h1
  · rw [show (2 : ℤ) * (↑m - ↑n) = -(↑(2 * n + 1) - ↑(2 * m + 1)) from by push_cast; ring]
    exact dvd_neg.mpr (dvd_sub (by exact_mod_cast Nat.gcd_dvd_right _ _)
      (by exact_mod_cast Nat.gcd_dvd_left _ _))
\end{lstlisting}

\subsection*{Step 2: Extract sub-problems with \extractdecls}

We then run \extractdecls to obtain each sub-problem as an independent, self-contained \lean snippet. The returned document map contains three entries: \texttt{gcd\_dvd\_two\_mul\_diff.h1}, \texttt{gcd\_dvd\_two\_mul\_diff.h2}, and the rewritten main lemma---each with the dependencies it needs to compile in isolation.

\subsection*{Step 3: Solve each sub-problem independently}

The first sub-problem is a straightforward divisibility fact that the agent solves on the first attempt:

\begin{lstlisting}[style=axledemo,language={}]
lemma gcd_dvd_two_mul_diff.h1 (m n : ℕ) :
    (2 * m + 1).gcd (2 * n + 1) ∣ 2 * m + 1 - (2 * n + 1) := by
  exact Nat.dvd_sub (Nat.gcd_dvd_left _ _) (Nat.gcd_dvd_right _ _)
\end{lstlisting}

The second is a simple arithmetic identity, so we save the LLM call and hand it directly to \repairproofs:

\begin{lstlisting}[style=axledemo,language=Python]
repaired = await client.repair_proofs(
    content=h2_sorry,
    names=["gcd_dvd_two_mul_diff.h2"],
    environment=ENVIRONMENT,
)
\end{lstlisting}

The tool closes the goal with a single terminal tactic:

\begin{lstlisting}[style=axledemo,language={}]
lemma gcd_dvd_two_mul_diff.h2 (m n : ℕ)
    (h1 : (2 * m + 1).gcd (2 * n + 1) ∣ 2 * m + 1 - (2 * n + 1)) :
    2 * m + 1 - (2 * n + 1) = 2 * (m - n) := by grind
\end{lstlisting}

\subsection*{Step 4: Reassemble with \merge}

We \merge the two solved sub-problems with the main decomposed proof. Declarations with matching types are reconciled by preferring the compiling version, and the surviving declarations are topologically ordered:

\begin{lstlisting}[style=axledemo,language=Python]
result = await client.merge(
    documents=[h1_proof, repaired.content, decomposed],
    environment=ENVIRONMENT,
)
final_proof = result.content
\end{lstlisting}

The merged output:

\begin{lstlisting}[style=axledemo,language={}]
import Mathlib

lemma gcd_dvd_two_mul_diff.h1 (m n : ℕ) :
    (2 * m + 1).gcd (2 * n + 1) ∣ 2 * m + 1 - (2 * n + 1) := by
  exact Nat.dvd_sub (Nat.gcd_dvd_left _ _) (Nat.gcd_dvd_right _ _)

lemma gcd_dvd_two_mul_diff.h2 (m n : ℕ)
    (h1 : (2 * m + 1).gcd (2 * n + 1) ∣ 2 * m + 1 - (2 * n + 1)) :
    2 * m + 1 - (2 * n + 1) = 2 * (m - n) := by grind

lemma gcd_dvd_two_mul_diff (m n : ℕ) :
    (Nat.gcd (2 * m + 1) (2 * n + 1) : ℤ) ∣ 2 * ((m : ℤ) - (n : ℤ)) := by
  have h1 : Nat.gcd (2 * m + 1) (2 * n + 1) ∣ (2 * m + 1) - (2 * n + 1) :=
    gcd_dvd_two_mul_diff.h1 m n
  have h2 : (2 * m + 1) - (2 * n + 1) = 2 * (m - n) :=
    gcd_dvd_two_mul_diff.h2 m n h1
  rw [h2] at h1
  rcases le_or_gt n m with hn | hn
  · rw [← Nat.cast_sub hn]; exact_mod_cast h1
  · rw [show (2 : ℤ) * (↑m - ↑n) = -(↑(2 * n + 1) - ↑(2 * m + 1)) from by push_cast; ring]
    exact dvd_neg.mpr (dvd_sub (by exact_mod_cast Nat.gcd_dvd_right _ _)
      (by exact_mod_cast Nat.gcd_dvd_left _ _))
\end{lstlisting}

\subsection*{Step 5: Final verification}

Finally, \verifyproof confirms that the assembled file actually proves the original formal statement:

\begin{lstlisting}[style=axledemo,language=Python]
result = await client.verify_proof(
    formal_statement=formal_statement,
    content=final_proof,
    environment=ENVIRONMENT,
)
# result.okay == True
\end{lstlisting}

The complete proof is now verified end-to-end. The pipeline used two generative-model calls (the initial attempt and the proof of \texttt{h1}); every other step was a mechanical invocation of \axle's tools. Decompose with \havetolemma and \extractdecls, solve sub-problems independently, reassemble with \merge, and check with \verifyproof---this is the basic shape of many agentic proving workflows \axle is designed to support.

\section{Strict Verification: Case Studies}
\label{app:verify-examples}

This appendix grounds the verification claims in concrete cases, organized into three groups:
\begin{enumerate}
    \item Two snippets that compile cleanly under \tool{check} but are rejected by \verifyproof's rule checks (\S\ref{app:case-vp-catches}).
    \item The residual verdict disagreements between \verifyproof and Comparator/SafeVerify on the $1{,}000$ paired requests of \S\ref{sec:eval-compare}---inter-tool differences that do not affect soundness on the user-named theorem (\S\ref{app:case-tool-diffs}).
    \item One adversarial kernel-bypass that \verifyproof accepts but a kernel-replay checker rejects (\S\ref{app:case-vp-misses}). This is the soundness gap for adversarial-input populations; we have not observed it in production traffic.
\end{enumerate}

\subsection{Cases \texttt{verify\_proof} Catches}
\label{app:case-vp-catches}

Both snippets below compile under plain \tool{check}; \verifyproof rejects each on a different rule.

\paragraph{Hidden axiom via a custom macro.} The snippet below defines a macro \texttt{totally\_safe\_decl} that, despite its name, expands to an \texttt{axiom} declaration---introducing \texttt{hello\_world~:~False}. A custom \texttt{fake\_simp} tactic then uses this axiom to ``prove'' \texttt{1 = 2}.

\begin{lstlisting}[style=axledemo,language={}]
syntax (name := totally_safe_decl) (priority := default + 1) declModifiers
  group("totally_safe_decl " declId ppIndent(declSig)) : command

@[macro «totally_safe_decl»] def expandTotallySafeDecl : Macro := fun stx =>
  let stx := stx.modifyArg 1 fun stx =>
    let stx := stx.modifyArg 0 (mkAtomFrom · "axiom" (canonical := true))
    stx.setKind ``Parser.Command.axiom
  pure <| stx.setKind ``Parser.Command.declaration

totally_safe_decl hello_world : False

elab "fake_simp" : tactic => do
  evalTactic (← `(tactic| exact hello_world))

theorem main : 1 = 2 := by exfalso; fake_simp
\end{lstlisting}

\verifyproof rejects \texttt{main} on the axiom-whitelist check, with error message:
\begin{quote}
\texttt{In 'main': Axiom 'hello\_world' is not in the allowed set of standard axioms}
\end{quote}

This is not an entirely contrived pattern. Math~Inc's FormalQualBench encountered it in the wild: both Codex and OpenCode submitted solutions that pass a plain \texttt{lake build} but reach the goal through an elaborator-level workaround.\footnote{\url{https://www.math.inc/formalqualbench}} One \texttt{BanachStoneTheorem} submission built an axiom via concatenation (\texttt{"ax"~++~"iom"}) and elaborated it at command time, so the keyword never appears literally in the source; the injected axiom then discharged the goal. \verifyproof properly rejects this pattern.

\paragraph{Namespace shadowing the intended definition.} Another case redefines \texttt{sqrt} inside a local namespace, then opens that namespace before stating a theorem whose name and surface syntax match the intended \mathlib statement.

\begin{lstlisting}[style=axledemo,language={}]
namespace hack
def sqrt (r : ℝ) := r
end hack

section
open hack

theorem impossible : sqrt 4 = 4 := by rfl

end
\end{lstlisting}

The file compiles---\texttt{rfl} closes the goal because \texttt{hack.sqrt 4} unfolds to \texttt{4}. \verifyproof rejects it on the signature-match check, with error message:
\begin{quote}
\texttt{Theorem 'impossible' does not match expected signature: expected type $\sqrt{4} = 4$, got hack.sqrt 4 = 4}
\end{quote}

\subsection{Inter-Tool Differences Without Soundness Implications}
\label{app:case-tool-diffs}

On the $1{,}000$-request paired sample (\S\ref{sec:eval-compare}), the only residual verdict disagreement is the $7$ SafeVerify-stricter cases of Table~\ref{tab:cmp-agreement} (\verifyproof and Comparator accept; SafeVerify rejects). All $7$ follow the same pattern. The candidate places \texttt{private} helpers inside a \texttt{namespace} and uses them in the proof of a public theorem. \lean mangles each \texttt{private} declaration's internal name with a module path---\texttt{\_private.\textit{Module}.0.\textit{Namespace}.\textit{name}}---so the same source compiled into two different modules ends up with two different mangled names. SafeVerify performs a whole-environment comparison between the target's \texttt{.olean} and the submission's \texttt{.olean}, and reports each mangled private helper present in one olean but not the other as ``declaration not found in submission''. The public theorem itself (the one named by the formal statement) is identical in both oleans and would type-check, but SafeVerify's checks the entire environment and rejects the whole submission.

A minimal example:

\begin{lstlisting}[style=axledemo,language={}]
import Mathlib

namespace Paper

private def castForm (x : ℝ) : ℝ := x + 1

private lemma castForm_add (x y : ℝ) :
    castForm (x + y) = castForm x + castForm y - 1 := by
  unfold castForm; ring

-- the user-named theorem the formal statement asks us to prove
theorem main (x : ℝ) : castForm x = x + 1 := rfl

end Paper
\end{lstlisting}

In the target olean (compiled from the formal statement), \texttt{castForm} and \texttt{castForm\_add} are stored as \texttt{\_private.Target.0.Paper.castForm} and \texttt{\_private.Target.0.Paper.castForm\_add}. In the submission olean (compiled from the candidate), the same source produces \texttt{\_private.Submission.0.Paper.castForm} and \texttt{\_private.Submission.0.Paper.castForm\_add}. SafeVerify walks every declaration in the target olean and looks it up by name in the submission olean; the \texttt{\_private.Target.*} names have no counterpart on the submission side, and SafeVerify reports them as missing:

\begin{lstlisting}[style=axledemo,language={},breaklines=true,basicstyle=\ttfamily\scriptsize]
Found a problem in Submission.olean with declaration
  _private.Target.0.Paper.castForm:
  declaration not found in submission
Found a problem in Submission.olean with declaration
  _private.Target.0.Paper.castForm_add:
  declaration not found in submission
\end{lstlisting}

\verifyproof and Comparator only check for matching signatures on the named theorem (\texttt{Paper.main}) and accept. None of these cases reflect a soundness gap on the user-named theorem.

\subsection{Cases \texttt{verify\_proof} Does Not Catch}
\label{app:case-vp-misses}

\verifyproof assumes every declaration in the environment was added through the standard, kernel-checked path, and so does not catch inputs that smuggle declarations into the environment by other means. The canonical example, from lean4checker's adversarial test suite, installs a declaration directly through \texttt{Environment.addDeclCore} with kernel checking disabled:

\begin{lstlisting}[style=axledemo,language={}]
import Lean.Elab.Term

open Lean in
run_elab
  modifyEnv fun env => Id.run do
    let decl := .thmDecl { name := `false, levelParams := [],
                            type := .const ``False [], value := .const ``False [] }
    let .ok env := env.addDeclCore (doCheck := false) 0 decl none |
      let _ : Inhabited Environment := ⟨env⟩
      unreachable!
    env
\end{lstlisting}

The result is an environment with a declaration \texttt{false~:~False}, never typechecked. \verifyproof accepts \texttt{false~:~False} as-is; a kernel-replay checker like lean4checker~\cite{lean4checker} re-runs the environment through the kernel from a clean state, detects the ill-formed declaration, and rejects.

Related kernel-bypass patterns work similarly: They evade the kernel's normal type-checking step at elaboration time and then use the unchecked declaration in the proof. The lean4checker, Comparator~\cite{comparator}, and SafeVerify~\cite{safeverify} test suites collect cases of this form.\footnote{\url{https://github.com/leanprover/lean4checker/tree/master/Lean4CheckerTests}, \url{https://github.com/leanprover/comparator/tree/master/tests/projects}, \url{https://github.com/GasStationManager/SafeVerify/tree/main/SafeVerifyTest}.} None of these patterns arise in the cooperating-client population \verifyproof targets; workloads verifying adversarial or external code should layer a strict checker on top. Readers interested in the broader landscape of \leanfour proof verification can consult the \lean reference.\footnote{\url{https://lean-lang.org/doc/reference/latest/ValidatingProofs/}}

\section{Resources}
\label{app:resources}

\ifblind
\noindent\textsc{[Resources table omitted under blind review. See published version for service URL, documentation, package installation, MCP endpoint, example notebook, release announcement, discussion forum, and support contact.]}
\else
\begin{tabularx}{\linewidth}{@{}lX@{}}
\toprule
\textbf{Resource} & \textbf{URL or command} \\
\midrule
Service                 & \url{https://axle.axiommath.ai} \\
Documentation           & \url{https://axle.axiommath.ai/v1/docs/} \\
Python SDK and CLI      & \texttt{pip install axiom-axle} \\
MCP server              & \texttt{pip install axiom-axle-mcp}, or hosted at \url{https://mcp.axiommath.ai} \\
Example notebook        & \url{https://colab.research.google.com/github/AxiomMath/axiom-lean-engine/blob/main/examples/starting_demo.ipynb} \\
Release announcement    & \url{https://axiommath.ai/territory/releasing-axle} \\
Discussion              & Axiom Lean Engine topic on the Lean Zulip \\
Email support           & \texttt{axle@axiommath.ai} \\
\bottomrule
\end{tabularx}
\fi

\end{document}